\documentstyle[12pt]{article}
\newcommand{\be}{\begin{equation}}
\newcommand{\ee}{\end{equation}}
\newcommand{\beq}{\begin{eqnarray}}
\newcommand{\eeq}{\end{eqnarray}}
\begin{document}
\def\a{\alpha}
\def\b{\beta}
\def\g{\gamma}
\def\G{\Gamma}
\def\d{\delta}
\def\e{\epsilon}
\def\z{\zeta}
\def\h{\eta}
\def\th{\theta}
\def\k{\kappa}
\def\l{\lambda}
\def\L{\Lambda}
\def\m{\mu}
\def\n{\nu}
\def\x{\xi}
\def\X{\Xi}
\def\p{\pi}
\def\P{\Pi}
\def\r{\rho}
\def\s{\sigma}
\def\S{\Sigma}
\def\t{\tau}
\def\f{\phi}
\def\F{\Phi}
\def\c{\chi}
\def\w{\omega}
\def\W{\Omega}
\def\costz{\frac{1}{\pi\a'}}
\def\intz{\int d^2z\;}
\def\zb{\bar{z}}
\def\de{\partial}
\def\deb{\bar{\partial}}
\def\cost{\frac{1}{4\pi\a'}}
\def\Ab{\bar{A}}
\def\ab{\bar{\a}}
\def\xt{\tilde{x}}
\def\yt{\tilde{y}}
\def\zt{\tilde{z}}
\def\xh{\hat{x}}
\def\xx{\vec{X}}
\def\Xt{\tilde{X}}
\def\gh{\hat{g}}
\def\Xh{\hat{X}}
\def\Pt{\tilde{P}}
\def\Ph{\hat{P}}
\def\et{\tilde{e}}
\def\eh{\hat{e}}
\def\Et{\tilde{E}}
\def\Eh{\hat{E}}
\def\Gt{\tilde{G}}
\def\Bt{\tilde{B}}
\def\ft{\tilde{\phi}}
\def\fb{\bar{\phi}}
\def\Qt{\tilde{Q}}
\def\tht{\tilde{\theta}}
\def\Lb{\bar{L}}
\def\bra{\langle}
\def\ra{\rightarrow}
\def\ket{\rangle}
\def\tr{{\rm Tr}}
\def\diag{{\rm diag}}
\def\NP{{\em Nucl. Phys.\/}}
\def\PL{{\em Phys. Lett.\/}}
\def\PR{{\em Phys. Rev.\/}}
\def\PRep{{\em Phys. Rep.\/}}
\def\RMP{{\em Rev. Mod. Phys.\/}}
\def\MPL{{\em Mod. Phys. Lett.\/}}
\def\CMP{{\em Comm. Math. Phys.\/}}
\def\PRL{{\em Phys. Rev. Lett.\/}}
\def\NC{{\em Nuovo Cimento\/}}
\def\NCL{{\em Nuovo Cimento Letters\/}}
\def\EL{{\em Europhys. Lett.\/}}

\titlepage
\begin{flushright}
DFTT-02/95 \\
ROM2F/94/36 \\
\end{flushright}
\vspace{6ex}
\begin{center}
{\bf HOMOGENEOUS CONFORMAL STRING BACKGROUNDS} \\
\vspace{4ex}
M.~Gasperini${}^{(a)}$ and R. Ricci${}^{(b)}$ \\
\vspace{8mm}

${}^{(a)}$
{\em Dipartimento di Fisica Teorica, Universit\`a di Torino, \\
Via P. Giuria 1, 10125 Turin, Italy \\
and INFN, Sezione di Torino, Turin, Italy} \\
\skip 0.5cm
${}^{(b)}$
{\em Dipartimento di Fisica, Universit\`a di Roma ``Tor Vergata'', \\
Via della Ricerca Scientifica 1, 00133 Rome, Italy} \\
\vspace{5ex}
{\small {\bf  ABSTRACT}}
\end{center}
We present exact solutions characterised by Bianchi-type I, II, III, V,
VI${}_0$, VI${}_h$ four-dimensional metric, space-independent dilaton, and
vanishing torsion background, for the low energy string effective action
with zero central charge deficit. We show that, in such a context,
curvature singularities cannot be avoided, except for the trivial case of flat
spacetime and constant dilaton. We also provide a further example of the
failure of the standard prescription for connecting conformal string
backgrounds through duality
transformations associated to
non-semisimple, non-Abelian isometry groups.

\vskip 1.5 cm
\noindent
--------------------------------

To appear in {\bf Class. Quantum Grav.}
\vspace{10 mm}
\vfill
\begin{flushleft}
ROM2F/94/36 \\
September 1994 \end{flushleft}
\section{Introduction}
\label{sect-1}
It is well known that the request for
conformal invariance of the sigma-model action
for closed (super)string theory implies, at the tree level in the string
loop expansion parameter $e^\f$, and to lowest order in the inverse of
the string tension $ \a'$, the background field equations of motion
\cite{beta-eq} %
\beq
R_\m{}^\n + \nabla_\m \nabla^\n \f - \frac{1}{4} H_{\m\a\b}H^{\n\a\b} = 0,
\label{beta.1}\\
R - \nabla_\m \f \nabla^\m \f + 2 \nabla_\m \nabla^\m \f + V
-\frac{1}{12}
H_{\m\n\a} H^{\m\n\a} = 0, \label{beta.2}\\
\partial_\m (e^{-\f} \sqrt{\vert g \vert} H^{\m\a\b}) = 0.
\label{beta.3}
\eeq
Here $V$ is a constant, $\f$ is the dilaton field, $H_{\m\n\a} = 6
\de_{[\m}B_{\n\a]}$ is the field strength of the antisymmetric (torsion)
tensor $B_{\m\n} = -B_{\n\m}$, and the covariant derivatives are performed
with respect to the background metric $g_{\m\n}$. These equations can be
derived from the low energy ($D$-dimensional) string effective action
\be
S =
- \int d^D X \sqrt{\vert g \vert} e^{-\f} (R +
\partial_\mu \f \partial^\mu \f + V - \frac{1}{12} H_{\m\n\a}
H^{\m\n\a}).
\label{S-eff}
\ee

In this paper we present a general procedure to integrate the
equations (\ref{beta.1})-(\ref{beta.3})
for the case of spatially homogeneous metric,
space-independent dilaton, vanishing torsion ($H_{\m\n\a} = 0$) and critical
dimension ($V=0$). This last requirement does not exclude, of course, the
phenomenologically interesting case of $d = D-1=3$, provided one adds the
right number of ``spectator'' dimensions in order to compensate the central
charge deficit. For $d=3$, in particular, our procedure can be applied to
obtain exact solutions for anisotropic but homogeneous
backgrounds, whose metric can be classified of type I, II, III, V,
VI${}_0$, VI${}_h$ according to the Bianchi classification (see for
instance \cite{Ry-Sh,Landau,ZelNov}).
These solutions may prove useful to study the correct
implementation of the
duality symmetry between conformal string backgrounds in the
case of non-Abelian isometries \cite{QdlO} (see for instance
the discussion in \cite{EGRSV} of the particular Bianchi V model
reported in \cite{GRV}).

We recall that, in the hypothesis of spatial
homogeneity, the $d$-dimensional spatial
submanifold is invariant under the action of a $d$-parameter isometry group
(generated by the $d$ Killing vectors $\x_i{}^\a$,  $i = 1, 2, \ldots, d$), and
the
metric can be factorized (in a synchronous  frame \cite{Landau}) as
\be
g_{00} = 1, \quad g_{0\a} = 0, \quad
g_{\a\b}(t, \vec{X}) = e_\a{}^i(\vec{X}) \g_{ij}(t) e_\b{}^j(\vec{X}),
\label{g}
\ee
($\a,\b = 1, \ldots, d$ are world indices in the spatial
submanifold). All dependence on the spatial coordinates $X^\a$ is thus
contained in the ``spatial'' {\em vielbein\/} $e_\a{}^i$, whose
corresponding Ricci rotation coefficients
\be
C_{ij}{}^k = e^\a{}_i e^\b{}_j ( \de_\a e_\b{}^k - \de_\b e_\a{}^k)
\label{Cijk}
\ee
are constant and determined by the algebraic structure of the isometry
group as \cite{Ry-Sh,Landau,ZelNov}
\be
[\x_i, \x_j] = C_{ij}{}^k \x_k, \qquad \x_i = \x_i{}^\a \de_\a.
\label{Lie}
\ee

Under the additional hypothesis that the dilaton field be
space-independent,
the background equations (\ref{beta.1})-(\ref{beta.3}) can be reduced to
ordinary time-differential equations for the variables $\g_{ij}(t), \f(t)$.
The spatial dependence of $R_\m{}^\n$ and $\nabla_\m \nabla^\n \f$
is determined, indeed, by the
choice of the isometry group, and can be factored out and eliminated through
a projection on the spatial {\em vielbein\/} $e_\a{}^i$
\cite{Landau,ZelNov}:
\beq
&& R_\a{}^\b \ra R_i{}^j = e^\a{}_i R_\a{}^\b e_\b{}^j, \\
&& R_\a{}^0 \ra R_i{}^0 = e^\a{}_i R_\a{}^0 , \\
&& \nabla_\a \nabla^\b \f \ra \nabla_i \nabla^j \f =
        e^\a{}_i e_\b{}^j \nabla_\a \nabla^\b \f  , \\
&& \nabla_\a \nabla^0 \f \ra \nabla_i \nabla^0 \f =
        e^\a{}_i \nabla_\a \nabla^0 \f.
\eeq
In particular, if we restrict our analysis to an anisotropic but diagonal
matrix form for the invariant metric $\g_{ij}$,
\be
\g_{ij}(t) =- a_i^2(t) \d_{ij},
\label{diag-g}
\ee
the projection gives
\beq
&& R_0{}^0 = R_0{}^0(\g_{mn}), \\
&& R_i{}^j =  \left[R_i{}^i(\g_{mn}) + V_i(\g_{mn}, C_{mn}{}^r)
\right] \d_i{}^j, \\
&& R_i{}^0 = \frac{1}{2} (\d_i{}^k C_{lj}{}^l - C_{ji}{}^k)
        \dot{\g}_{kl} \g^{lj}, \\
&& \nabla_0 \nabla^0 \f = \ddot{\f}, \\
&& \nabla_i \nabla^j \f = \dot{\f} H_i \d_i{}^j, \\
&& \nabla_i \nabla^0 \f = 0,
\eeq
 (no sum over $i$; a dot denotes differentiation with respect to the
cosmic time $t$). Here $H_i = \dot{a}_i/a_i$, and $R_0{}^0(\g)$,
$R_i{}^j(\g)$  are the time and space components of the Ricci tensor for
the metric (\ref{diag-g}). The ``effective potentials'' $V_i(\g,C)$
(vanishing
for $C_{ij}{}^k=0$) represent the explicit contribution of the non-Abelian
part of the isometry group and are related to the Riemann curvature of the
spatial submanifold.

{}From the $(0,0)$ and $(i,i)$ part of the background eq.~(\ref{beta.1}) we
thus
obtain, respectively,
\beq
&& \sum_i (\dot{H}_i + H_i^2) - \ddot{\f} = 0,
\label{eq.1} \\
&& \dot{H}_i + H_i \sum_k H_k - H_i \dot{\f} - V_i = 0,
\label{eq.2}
\eeq
while the mixed components $(i,0)$ give the constraint
\be
\sum_{k=1}^d C_{ki}{}^k (H_i - H_k) = 0,
\label{constraint}
\ee
(no sum over $i$). The dilaton equation (\ref{beta.2}) moreover implies
\be
2 \ddot{\f} - \dot{\f}^2 + 2\dot{\f} \sum_k H_k + \sum_k V_k - (\sum_k
H_k)^2  - \sum_k H_k^2 -2\sum_k \dot{H}_k = 0
\label{dil->}
\ee
In the following section it will be shown that the above equations
(\ref{eq.1})-(\ref{dil->}) can be integrated exactly, provided the potential
functions $V_i(a_j)$ satisfy particular restrictions.

\section{General integration method for a class of homogeneous backgrounds}
\label{sect-2}
In order to integrate the equations (\ref{eq.1})-(\ref{dil->})
we shall try to extend to the more general
homogeneous case a procedure already successfully applied to
space-independent metric backgrounds even in the presence of string sources
\cite{PBB,infl-defl}, non-vanishing torsion \cite{dil-prod}
and a particular class of dilaton potentials \cite{PBB}.

We introduce, first of all, the rescaled dilaton $\fb$,
\be
\fb = \f - \frac{1}{2} \ln |\det(\g_{ij})| = \f - \sum_j \ln a_j,
\label{dil-bar}
\ee
which is exactly the duality-invariant variable defined in the
context of the  particular
``scale factor'' duality symmetry, for
space-independent cosmological metrics
\cite{V,Tse}.
In terms of this variable the equations (\ref{eq.1}), (\ref{eq.2}),
(\ref{dil->})
become respectively:
\beq
&& \ddot{\fb} - \sum_i H_i^2 = 0,
\label{e1} \\
&& \dot{H}_i - H_i \dot{\fb} - V_i = 0,
\label{e2} \\
&& \dot{\fb}^2 - 2\ddot{\fb} + \sum_i H_i^2 - \sum_i V_i = 0.
\label{e3}
\eeq
The combination of eqs.~(\ref{e1}) and (\ref{e3}) gives
\be
\dot{\fb}^2 - \sum_i H_i^2 - \sum_i V_i = 0.
\label{e4}
\ee
By differentiating the equation above and using (\ref{e1}), (\ref{e2}) to
eliminate $\ddot{\fb}$, $\dot{H}_i$, we get
\be
\sum_i (\dot{V}_i + 2 H_i V_i) = 0,
\label{e5}
\ee
which can be interpreted as a sort of covariant conservation equation for
the effective ``source density'' $\sum_i V_i$, following from the Bianchi
identities of the effective scalar-tensor theory.

We choose now eqs. (\ref{e2})-(\ref{e4}) as independent equations,
and we show
that they can be integrated exactly for all $V_i$ satisfying the condition
\be
V_i = k_i \sum_{j=1}^d V_j,
\label{cond}
\ee
where $k_i$ can be arbitrary real numbers.

By combining eqs.~(\ref{e3}), (\ref{e4}) we get in fact
\be
(e^{-\fb})\;\ddot{} = e^{-\fb} \sum_j V_j,
\label{ee1}
\ee
while eq.~(\ref{e2}), using (\ref{cond}), can be rewritten as
\be
(e^{-\fb} H_i)\dot{} = e^{-\fb} k_i \sum_j V_j.
\label{ee2}
\ee
If we substitute $t$ for a new dimensionless time-like variable $x$,
defined by
\be
\frac{1}{L} \frac{d x}{d t} = e^{-\fb} \sum_j V_j,
\label{x-def}
\ee
($L$ is an appropriate dimensional constant), eqs.~(\ref{ee1}) and
(\ref{ee2}) can
be integrated a first time to give
\beq
&& (e^{-\fb})' e^{-\fb} \sum_j V_j = \frac{(x + x_0)}{L^2},
\label{eee1} \\
&& \frac{a_i'}{a_i} e^{-\fb} \sum_j V_j = \frac{e^{\fb}}{L^2}\G_i,
\label{eee2}
\eeq
where
\be
\G_i = k_i x + x_i
\label{gamma-i}
\ee
($x_i$, $x_0$ are integration constants, and a prime denotes
differentiation with respect to $x$).
Moreover, using eqs.~(\ref{cond}) and
(\ref{eee2}), the identity (\ref{e5}) can be written as
\be
\sum_j V'_j = - \frac{e^{2\fb}}{L^2} \sum_j (\G_j^2)'.
\label{eee3}
\ee
By adding eqs.~(\ref{eee1}), (\ref{eee3}), and integrating, we thus obtain the
important constraint
\be
L^2 e^{-2\fb} \sum_j V_j = \b + (x + x_0)^2 - \sum_j \G_j^2,
\label{important}
\ee
which allows the separation of variables in eqs.~(\ref{eee1}), (\ref{eee2})
and which, as we
shall see, ultimately defines the range of validity of our solution with
respect to
the $x$ coordinate ($\b$ is an integration constant).

The constant $\b$ appearing in eq.~(\ref{important})
is not arbitrary. Indeed, out of the three
independent equations (\ref{e2})-(\ref{e4}) we have used, up to now, only
eq.~(\ref{e2}) and a linear combination of eqs.~(\ref{e3}) and (\ref{e4}). We
still have the freedom to impose that eq.~(\ref{e4}) be also separately
satisfied by the result of our first integration, eqs. (\ref{eee1}),
(\ref{eee2}).
By computing  $\dot{\fb}$ and $H_i$ from eqs.~(\ref{x-def})-(\ref{eee2}), and
inserting their  values into eq.~(\ref{e4}), we find that this last equation is
identically  satisfied, and compatible with eq.~(\ref{important}),
if and only if $\b = 0$. Using eq. (\ref{important}) (with $\b=0$)
the system of coupled differential equations (\ref{eee1}),
(\ref{eee2}) can be consistently reduced to quadratures, and we are
eventually led to
\beq
\fb' = -\frac{x+x_0}{D(x)},
\label{q1} \\
\frac{a_i'}{a_i} = \frac{\G_i}{D(x)},
\label{q2}
\eeq
where the quadratic form $D(x)$ must satisfy the condition
\be
D(x) \equiv  (x+ x_0)^2 - \sum_i \G_i^2 = L^2 e^{-2\fb} \sum_j V_j.
\label{D}
\ee

Our background equations can thus be integrated exactly for all homogeneous
metrics satisfying eq.~(\ref{cond}), and the solution is valid for the range of
$x$ compatible with the constraint (\ref{D}). Moreover,
the allowed values of the constant
``charges'' $k_i$, and of the integration constants $x_i$, are further
restricted by the mixed components of the background equations,
$R_i{}^0=0$. The insertion of eq.~(\ref{q2}) into eq.~(\ref{constraint})
gives in fact the additional constraints on the solution
\be
\sum_k C_{ki}{}^k (k_i - k_k) = 0, \quad \sum_k C_{ki}{}^k (x_i - x_k) = 0
\label{additional}
\ee
(no sum over $i$).

We finally note that our integration procedure obviously applies also to
the trivial case $V_i=0$ (Abelian isometry group of spatial translations).
In this case, however, there is no need to introduce a new time variable
and from eqs.~(\ref{ee1}), (\ref{ee2}) we obtain directly
\be
e^{\fb} = \frac{L}{c_0 t + t_0}, \qquad H_i = \frac{c_i}{c_0 t + t_0}
\label{sol-I}
\ee
where $c_i$, $c_0$, $t_0$, $L$ are integration constants, related by the
condition
\be
c_0^2 = \sum_i c_i^2,
\label{sol-Ibis}
\ee
which is required
in order to satisfy separately also eq.~(\ref{e4}). One thus recovers the
well-known ``Kasner-like'' anisotropic background \cite{V,Mueller},
first derived in
the context of the Brans-Dicke solutions in vacuum \cite{RF}.

In the following section we shall apply the integration procedure
just outlined to the
case of homogeneous cosmological backgrounds in $d=3$ spatial dimensions.

\section{Bianchi-type solutions and curvature singularities}
\label{sect-3}
Homogeneous manifolds with $d=3$ spatial dimensions can be
classified in nine different Bianchi
types \cite{Ry-Sh,Landau,ZelNov}, according to the structure of
their isometry groups.
By considering the explicit form of the potential functions $V_i(a_j)$ for
the various metric types (see for instance \cite{Chauvet}),
one finds that the
conditions of applicability of our integration procedure are met for
Bianchi types I, II and VI${}_h$ (in the notations of Ref.~\cite{Ry-Sh}).
This last case includes Bianchi types III, V and VI${}_0$, corresponding to
$h = 0$, $1$ and $-1$ respectively.

Bianchi I type is characterised by an Abelian isometry group, $V_i=0$, and in
this case the integration of eqs.~(\ref{sol-I}) leads to the previously quoted
solution \cite{V,Mueller,RF}. For a Bianchi II metric there is only one
non-vanishing structure constant,
\be
C_{31}{}^2 = 1 = -C_{13}{}^2
\ee
and eq.~(\ref{cond}) is satisfied with
\be
{k_i} = (-1, 1, 1), \quad \,\, L^2\sum_j V_j = \frac{a_1^2}{2 a_2^2 a_3^2} \geq
0
\label{k-II}
\ee
For Bianchi VI${}_h$ the structure constants are
\be
C_{21}{}^2 = 1~~~~,~~~~~  C_{31}{}^3 = h
\ee
and eq.~(\ref{cond}) is satisfied with
\be
{k_i} = \frac{1}{2(1 + h + h^2)} (1 + h^2, 1 + h, h + h^2), \quad
L^2\sum_j V_j = \frac{2}{a_1^2}(1 + h + h^2) \geq 0
\label{k-VIh}
\ee
One can easily verify that the constants $k_i$ of the above Bianchi models
also automatically satisfy the constraint (\ref{additional}).

In the case of Bianchi II and Bianchi VI${}_h$ metric, the general form of
the background solution is thus provided by the explicit integration of
eqs.~(\ref{q1}), (\ref{q2}).
By calling $x_\pm$ the two real zeros of $D(x)$
(the case of complex roots, and of real but coincident roots $x_+=x_-$,
 will be discussed below) we obtain
\beq
&& \frac{a_i}{a_{i0}} =
\vert (x - x_+) (x - x_-) \vert^{\frac{k_i}{2\a}}
\biggl
\vert\frac{x - x_+}{x - x_-} \biggr \vert^{\frac{\a_i}{2}},
\label{sol.1} \\
&& e^{\fb} = e^{\fb_0} \vert (x - x_+) (x - x_-) \vert^{-\frac{1}{2\a}} \biggl
\vert\frac{x - x_+}{x - x_-} \biggr \vert^{-\frac{1}{2}\sum_i k_i \a_i},
\label{sol.2}
\eeq
where $a_{i0}$, $\fb_0$ are integration constants, and
\beq
&& \a = 1 - \sum_i k_i^2,
\label{a.1}\\
&& \a_i = \frac{\a x_i + k_i(\sum_j k_j x_j - x_0)}
{\a\sqrt{(\sum_j k_j x_j - x_0)^2 + \a(\sum_j x_j^2 - x_0^2)}},
 \label{a.2}\\
&& \sum \a_i k_i = \frac{\sum_i k_i x_i - x_0 \sum_i k_i^2}
{\a\sqrt{(\sum_j k_j x_j - x_0)^2 + \a(\sum_j x_j^2 - x_0^2)}},
 \label{a.3}\\
&& x_\pm = \frac{1}{\a} \left(\sum_j k_j x_j - x_0 \pm
\sqrt{(\sum_j k_j x_j - x_0)^2 + \a(\sum_j x_j^2 - x_0^2)} \right).
 \label{a.4}
\eeq

The coefficients $k_i$ are given by eqs.~(\ref{k-II}) and
(\ref{k-VIh}) for Bianchi
types II and VI${}_h$ respectively, and the integration constants $x_i$
must satisfy the constraint (\ref{additional}),
which for the Bianchi VI${}_h$ type reads explicitly
\be
(1 + h) x_1 = x_2 + h x_3.
\label{add-VIh}
\ee
Further restrictions on the solutions follow from eq. (\ref{D}), which in the
Bianchi II case imposes a relation among the integration constants
$a_{i0}$, $\fb_0$, and which in the Bianchi  VI${}_h$ case also defines the
allowed range of $h$, for any given choice of the integration constants. It is
interesting to note that for $h=1$ (Bianchi V), a possible choice is the
particular case $x_1 = x_2 = x_3$, which leads to an isotropic homogeneous
solution
with $a_1 = a_2 = a_3$. Such solution
represents a Friedman-Robertson-Walker conformal string background with
constant (negative) spatial curvature, while the isotropic version of
the Bianchi I solution (\ref{sol-I}) represents the corresponding background
with vanishing spatial curvature.

The temporal range of validity of the solution (\ref{sol.1}), (\ref{sol.2})
is also determined by eq.~(\ref{D}), which implies
\be
{\rm sign}(D) = {\rm sign}(\sum_j V_j) \geq 0.
\label{sign}
\ee
For a Bianchi II metric we have $\a < 0$ (see eq.~(\ref{k-II})),
and the solution is thus defined in the limited range
\be
x_- < x < x_+.
\ee
For a Bianchi VI${}_h$ metric we must treat separately the particular case
$h=-1$ (Bianchi VI${}_0$), for which $\a=0$ and the quadratic form $D(x)$
degenerates in a line which crosses the $x$ axis at
\be
x = x_c = \frac{\sum_j x_j^2 - x_0^2}{2(x_0 - x_1)}.
\ee
The solution is defined, in this case, on the half-line $x > x_c$.
For all other values of $h$ we have $\a > 0$ and the solution is
characterised by two branches, defined on the two half-lines
\be
x < x_-, \qquad x > x_+.
\ee

In correspondence of the two roots of $D(x)$ both $H_i$ and $\exp(\fb)$
diverge, and the background solutions run into a singularity of both the
curvature and the effective string coupling constant.
A similar singularity occurs for the Bianchi I solution (\ref{sol-I}),
which is
characterised by two branches, defined on the two half-lines
\be
t <- t_0/c_0, \qquad t >- t_0 /c_0
\ee
and separated by a curvature singularity at $ t =- t_0/c_0$.
Such singularities cannot be avoided in the context of the low energy
string effective action considered here, except for the trivial case of
flat spacetime and constant dilaton solution.

Indeed, necessary conditions to prevent divergences of the curvature and
dilaton background turn out to be
1) the absence of real zeros of $D(x)$ or 2) the coincidence of the two
real zeros of $D(x)$ among themselves and with the zeros of the two
numerators at the right-hand-side of eqs.~(\ref{q1}), (\ref{q2}),
namely $x+x_0 = 0 = k_ix + x_i$, where $D(x)=0$.

If the quadratic form $D(x)$
\be
D(x) = (x + x_0)^2 - \sum_i(k_ix + x_i)^2 = \a (x - x_+)(x- x_-)
\ee
has no real zeros, however, it must be always negative.
Therefore, the first  requirement cannot be satisfied
neither by Bianchi II nor by  Bianchi VI${}_h$ solutions, as it
would be in contradiction with the  condition (\ref{sign}).
In the Bianchi I case the first requirement could be
satisfied by the choice $c_0 = 0$, but this implies that all the constants
$c_i$ are vanishing, namely that the solution is trivial
(see eqs. (\ref{sol-I}), (\ref{sol-Ibis})).

The second requirement can be met by choosing the integration constants
$x_i$ in such a way that the two real roots of $D(x)$ coincide with $x_0$,
namely for
\be
x_i= k_i x_0,~~~~ x_+=x_- =-x_0,~~~~ D(x)= \a(x+x_0)^2
\ee
In this case, however, the Bianchi II and Bianchi VI${}_0$ solutions are
consistently defined only on a point (where $\sum_j V_j=0$), according to
eq.~(\ref{sign}). For a Bianchi VI${}_h$ metric ($h \neq -1$), on the contrary,
the range of validity is non-trivial,
and the solution is defined by the equations
\be
\fb' = -\frac{1}{\a(x+x_0)}, \qquad \frac{a_i'}{a_i} = \frac{k_i}{\a(x +
x_0)}, \qquad \a = 1 - \sum_i k_i^2,
\ee
with the coefficients $k_i$ of eq.~(\ref{k-VIh}). Their integration gives
\be
\fb = \fb_0 + \ln \vert x+ x_0 \vert^{-\frac{1}{\a}}, \qquad
a_i = a_{i0} \vert x+ x_0 \vert^{\frac{k_i}{\a}},
\ee
where $\fb_0$ and $a_{i0}$ are integration constants.

This solution, however, is only valid for the set of values
of $\fb_0$, $a_{i0}$ and $h$ satisfying the constraint (\ref{D}). As a
consequence, its dynamical content is trivial, as one can easily check by
noting first of all that the dilaton background is constant (according to the
definition (\ref{dil-bar})),
\be
\f = \fb + \sum_j \ln a_j = \fb_0 + \sum_i \ln a_{i0} = const,
\label{d}
\ee
since $\sum_i k_i =1$. Moreover, choose for instance the integration
constants in such a way that the scale factors, when expressed in cosmic
time according to eq. (\ref{x-def}), are given by
\be
a_i(t)= |t|^{\beta_i}, \qquad
{\beta_i} = (1, \frac{1+h}{1+h^2}, \frac{h(1+h)}{1+h^2})
\label{ab}
\ee
and the full Bianchi VI${}_h$ metric ($h \neq -1$) takes the form
\be
g_{\mu\nu}(\vec{X},t) = \diag (1,-t^2,-t^{2 \b_2}e^{-2X},
-t^{2\b_3}e^{-2hX}).
\label{ac}
\ee
The constraint (\ref{D}) implies then a condition on $h$ which is only
satisfied, for real values of the parameter, by $h=0$ and $1$
(see the Appendix). In both
cases, the solution (\ref{ac}) is identically Ricci flat and Riemann flat
(see the Appendix), showing that  also
the metric background is
trivial.
\section{Conclusion}
\label{sect-4}
In this paper we have presented a procedure for obtaining homogeneous
background solutions for the low energy string effective action. Such
solutions are characterized by a spatial, generally non-Abelian transitive
isometry group, and may be
useful for investigating possible extensions of the $O(d,d)$ covariance
(see \cite{Meis} and references therein) associated
to backgrounds with Abelian translational symmetry.
Moreover, in $d
=3$ spatial dimensions such solutions correspond to homogeneous Bianchi
type models, which may be of some phenomenological interest for
applications to a very early cosmological regime with non-vanishing
anisotropy and time-varying dilaton field. The explicit form of the
metric and dilaton field, for the particular case of Bianchi I, II, III,
V and VI${}_0$ models, is given explicitly in {\bf Table I}.

The solutions reported in the table refer to the case in which the zeros
of $D(x)$ are real and both different from the zeros of $\Gamma_i$ and
of $x+x_0$ (otherwise the dilaton is constant, and the metric globally
flat up to reparametrizations).
The solutions (except those of the Bianchi II and Bianchi VI${}_0$
type) in  general exhibit
two  branches, characterised respectively by a final and an initial
curvature  singularity (a similar behaviour is also typical of Bianchi I
backgrounds with nontrivial torsion, $H_{\mu\nu\a} \neq 0$, as recently
discussed in  \cite{Cop}).
The singularities cannot be avoided in this context, but
they could be eventually cured by higher order corrections in $\alpha^
{\prime}$ and in the string loop expansion parameter, which become
important when approaching the high curvature, strong coupling regime
surrounding the singularity.

We finally note that the trivial solution (\ref{d})-(\ref{ac}) suggests
particularly simple examples of conformal backgrounds suitable for
performing duality transformations with respect to a non-Abelian isometry
group. The case of a Bianchi V metric ($h=1$) was already discussed in
\cite{GRV}. The Bianchi III case ($h=0$),
\be
g_{\mu\nu} = \diag (1,-t^2,-t^2 e^{-2X},-1), \qquad
\phi = const,
\label{metrica}
\ee
also corresponds to a non-semisimple, non-Abelian group of isometries,
with $C_{21}{}^2 =1$ as the only non-vanishing structure constant. By
following the standard prescriptions \cite{QdlO,GRV}, the non-Abelian
duality transformations applied to eq.(\ref{metrica}) lead to a dual metric
which  is still diagonal,
\be
\tilde g_{\mu\nu} = \diag (1, -\frac{t^2}{\Delta}, -\frac{t^2}{\Delta},
-1), \qquad \Delta = t^4+Y^2,
\label{tilde}
\ee
but also to a non-vanishing torsion and a non-trivial dilaton field,
\be
\tilde B_{12} = \frac {Y}{\Delta} = -\tilde B_{21}, \qquad
\tilde \phi =-\ln \Delta + const.
\label{tor}
\ee
Since
\be
e^{-\tilde \phi}\sqrt{|\tilde g|} =t^2, \quad \tilde H^{201}
=-\frac{4Y}{t}, \quad \tilde H^{301}=0=\tilde H^{321},
\ee
it follows that the dual background is not conformal, as one can easily
check by noting for instance that the component $\alpha=0$, $\beta=1$ of
eq.(\ref{beta.3}) is not satisfied,
\be
\partial_2\left ( e^{-\tilde \phi}\sqrt{|\tilde g|} \tilde H^{201}\right)
=-4t \neq 0.
\ee

By following the same procedure as in \cite{GRV} one can show, in
particular, that no possible choice of the transformed dilaton can
restore conformal invariance for the dual background $\{\tilde g ,
\tilde B\}$ defined in (\ref{tilde}), (\ref{tor}).
This confirms a recent analysis
\cite{louis} showing that, in the case of non-semisimple groups, an
additional anomaly cancellation condition is to be imposed for the
consistency of non-abelian duality.
\section{Aknowledgements}
We are very grateful to G.~Veneziano for many discussions and helpful
suggestions. We also wish to thank the Theory Division at CERN
for its warm hospitality and financial support during part of this work.

\section{Appendix}
In order to compute the allowed values of $h$ for the particular solution
(\ref{d}), (\ref{ab}),
\be
a_i(t)= t^{\beta_i}, \quad
{\beta_i} = (1, \frac{1+h}{1+h^2}, \frac{h(1+h)}{1+h^2}), \quad
h\neq -1,
\ee
\be
\f = c = const, \qquad \fb = c -\sum_i \b_i \ln t,
\ee
we rewrite it in terms of the $x$ coordinate. By recalling that, for a Bianchi
type  VI${}_h$,
\be
V_i= {1\over t^2}(1+h^2, {1+h}, {h+h^2}),
\ee
we obtain from eq. (\ref{x-def})
\be
x+x_0= {2(1+h+h^2) \over \sum_k \b_k -1} t^{\sum_k
\b_k -1} e^{-c}
\ee
(we have put $L=1$ for simplicity). It follows that
\be
a_i= a_{i0}(x+x_0)^{\b_i\over \sum_k \b_k -1}, \qquad
e^{-\fb}=e^{-\fb_0} (x+x_0)^{\sum_i\b_i\over \sum_k \b_k -1},
\ee
where
\be
a_{i0}=\left[\sum_k \b_k -1 \over
e^{-c}2(1+h+h^2)\right]^{\b_i\over \sum_k \b_k -1}, \quad
e^{-\fb_0}= e^{-c} \left[\sum_k \b_k -1 \over
e^{-c}2(1+h+h^2)\right]^{\sum_i\b_i\over \sum_k \b_k -1}.
\ee
By inserting these values into the constraint (\ref{D}) we thus obtain
the condition
\be
\a={2\over a_{10}^{2}}(1+h+h^2) e^{-2\fb_0}
\ee
which reads explicitly
\be
{(1+h)^4(1+h+h^2)\over (1+h^2)^2} =
2(1+h+h^2) - (1+h+2h^2+h^3+h^4)
\ee
and which, for $h$ real, is only satisfied by $h=0,1$ ($h=-1$ is also
allowed, but this value is to be excluded for the particular solution we
are considering, see Sect. 3). For these two values of $h$ the full
Bianchi metric (\ref{ac}) is identically Ricci flat,
\be
R_1\,^1={(h-1)h(2+h+h^2) \over (1+h^2)t^2} \equiv 0, \qquad
R_0\,^0 \equiv 0,
\ee
\be
R_2\,^2 = {1+h\over 1+h^2}R_1\,^1, \qquad
R_3\,^3= {h(1+h)\over 1+h^2}R_1\,^1
\ee
as required by a solution of the background field equations with
constant dilaton. However, the spacetime manifold is also globally flat,
since for the metric (\ref{ac}) all the components of the Riemann
tensor are proportional to $h(h-1)$, and thus identically vanishing for
$h=0,1$. In particular,
\be
R_{1212}={h(h-1)\over 1+h^2}t^{2(1+h)\over 1+h^2} e^{-2X},
\ee
\be
R_{1220}=-R_{1212}t^{-1},
\ee
\be
R_{1313}=(1+h+h^2) R_{1212} t^{2(h^2-1)\over 1+h^2} e^{2X(1-h)},
\ee
\be
R_{1330}=(1+h+h^2)^{-1} R_{1313} t^{-1},
\ee
\be
R_{2020}={1+h\over 1+h^2}R_{1220} t^{-1},
\ee
\be
R_{3030}={1+h\over 1+h^2}R_{1330} t^{-1},
\ee
\be
R_{2323}={h(2+h+h^2)\over 1+h}R_{3030} t^{-2(1+h)\over 1+h^2}e^{-2X}.
\ee

\vfill\eject

\vfill\eject

\section{Table caption}
\vskip 2 cm
Explicit form of the metric and dilaton field in $d=3$ for Bianchi types
I, II, III,
V and VI${}_0$ ($\alpha_i, x_{\pm}$ and $x_c$ are defined respectively
by eqs.(51), (53) and (57)). The solutions reported here refer to the
non-trivial case in which the zeros of $D(x)$ are real and different
from $-x_0$ and from the zeros of $\Gamma_i$. The range of validity of
such solutions is discussed in Section 3.

\vfill\eject
\centerline{\bf Table 1}

\noindent
------------------------------------------------------------------------
--------------------------

\noindent
{\bf Bianchi I}
$$
ds^2= dt^2-a_1^2dX^2-a_2^2dY^2-a_3^2dZ^2
$$
$$
a_i = a_{i0} (t_0+c_0 t)^{c_i/c_0}, ~~~~~~~~~~~~~~~~~
\sum_i c_i^2 = c_0^2, ~~~~~~~~~~~~~~~~ i=1,2,3
$$
$$
e^{\fb} = {e^\phi \over a_1 a_2 a_3} = L (c_0 t+t_0)^{-1}
$$
------------------------------------------------------------------------
--------------------------

\noindent
{\bf Bianchi II}
$$
ds^2= dt^2-a_1^2dX^2-a_2^2(dY-XdZ)^2-a_3^2 dZ^2
$$
$$
a_i=a_{i0}
\vert (x - x_+) (x - x_-) \vert^{\frac{-k_i}{4}}
\biggl
\vert\frac{x - x_+}{x - x_-} \biggr \vert^{\frac{\a_i}{2}},
{}~~~~~~~~~~~~~ k_i=(-1,1,1)
$$
$$
e^{\fb} = e^{\fb_0} \vert (x - x_+) (x - x_-) \vert^{\frac{1}{4}} \biggl
\vert\frac{x - x_+}{x - x_-} \biggr \vert^{\frac{1}{2}(\a_1 -\a_2-\a_3)}
$$
------------------------------------------------------------------------
--------------------------

\noindent
{\bf Bianchi III}
$$
ds^2= dt^2-a_1^2dX^2-a_2^2 e^{-2X}dY^2-a_3^2dZ^2
$$
$$
a_i=a_{i0}
\vert (x - x_+) (x - x_-) \vert^{k_i}
\biggl
\vert\frac{x - x_+}{x - x_-} \biggr \vert^{\frac{\a_i}{2}},
{}~~~~~~~~~~~~~ k_i=({1\over 2},{1\over 2},0)
$$
$$
e^{\fb} = e^{\fb_0} \vert (x - x_+) (x - x_-) \vert^{-1} \biggl
\vert\frac{x - x_+}{x - x_-} \biggr \vert^{-\frac{1}{4}(\a_1 +\a_2)}
$$
------------------------------------------------------------------------
--------------------------

\noindent
{\bf Bianchi V}
$$
ds^2= dt^2-a_1^2dX^2-a_2^2 e^{-2X}dY^2-a_3^2 e^{-2X}dZ^2
$$
$$
a_i=a_{i0}
\vert (x - x_+) (x - x_-) \vert^{\frac{3 k_i}{4}}
\biggl
\vert\frac{x - x_+}{x - x_-} \biggr \vert^{\frac{\a_i}{2}},
{}~~~~~~~~~~~~~ k_i=({1\over 3},{1\over 3},{1\over 3})
$$
$$
e^{\fb} = e^{\fb_0} \vert (x - x_+) (x - x_-) \vert^{-\frac{3}{4}}
\biggl
\vert\frac{x - x_+}{x - x_-} \biggr \vert^{-\frac{1}{6}(\a_1 +\a_2
+\a_3)}
$$
------------------------------------------------------------------------
--------------------------

\noindent
{\bf Bianchi VI$_0$}
$$
ds^2= dt^2-a_1^2dX^2-a_2^2 e^{-2X}dY^2-a_3^2 e^{2X}dZ^2
$$
$$
a_1=a_{10}(x-x_c)^{\frac{x_c+x_1}{2(x_0-x_1)}}e^{\frac{x}{2(x_0-x_1)}}
$$
$$
a_2=a_{20}(x-x_c)^{\frac{x_2}{2(x_0-x_1)}}
$$
$$
a_3=a_{30}(x-x_c)^{\frac{x_3}{2(x_0-x_1)}}, ~~~~~~~~~~~~~x_2=x_3
$$
$$
e^{\fb} = e^{\fb_0}
(x-x_c)^{-\frac{x_c+x_0}{2(x_0-x_1)}}e^{-\frac{x}{2(x_0-x_1)}}
$$
------------------------------------------------------------------------
--------------------------


\begin{thebibliography}{99}
%
\bibitem{beta-eq} Fradkin E S and Tseytlin A A 1985
\NP\ {\bf B261} 1; \\
Callan C G, Friedan D, Martinec E J and Perry M J 1985 {\em
Nucl. Phys.\/} {\bf B262} 593
%
\bibitem{Ry-Sh} Ryan M P and Shepley L C 1975 {\em Homogeneous
Relativistic Cosmologies\/} (Princeton: Princeton University Press)
%
\bibitem{Landau} Landau L D and  Lifshits E M 1987 {\em The Classical Theory of
Fields\/} (Pergamon Press)
%
\bibitem{ZelNov} Zel'dovich Y B and Novikov I D 1983 {\em Relativistic
Astrophysics\/} (Chicago: Chicago Press)
%
\bibitem{QdlO} de la Ossa X C and Quevedo F 1993
{\em Nucl. Phys.\/} {\bf B403} 377
%
\bibitem{EGRSV} Elitzur S, Giveon A, Rabinovici E, Schwimmer A and
Veneziano G 1994  {\em Preprint\/} CERN-TH.7414/94

\bibitem{GRV} Gasperini M, Ricci R and Veneziano G, 1993 \PL\
{\bf B319} 438
%
\bibitem{PBB} Gasperini M and Veneziano G 1993
{\em Astroparticle Phys.\/}
{\bf 1}  317
%
\bibitem{infl-defl} Gasperini M and Veneziano G 1993 \MPL\
{\bf A8} 3701
%
\bibitem{dil-prod} Gasperini M and Veneziano G 1994
\PR\ {\bf D50} 2519
%
\bibitem{V} Veneziano G 1991 {\em Phys. Lett.\/} {\bf B265} 287
%
\bibitem{Tse} Tseytlin A A 1991 {\em Mod. Phys. Lett.\/} {\bf A6} 1721
%
\bibitem{Mueller} Mueller M 1990 \NP\ {\bf B337} 37
%
\bibitem{RF} Ruban V A and Finkelstein A M 1972 \NCL\
{\bf 5} 289
%
\bibitem{Chauvet} Chauvet P, Cervantes-Cota J and  N\'u\~{n}ez-Y\'epez H
N 1992 {\em Class. Quantum Grav.\/} {\bf 9} 1923
%
\bibitem{Meis} Meissner K A and Veneziano G 1991 \PL\ {\bf B267} 33; \\
Sen A 1991 \PL\ {\bf B271} 295; \\
Hassan S F and Sen A 1992 \NP\ {\bf B375} 103;\\
Gasperini M and Veneziano G 1992 \PL\ {\bf B277} 256;\\
Giveon A, Porrati M and Rabinovici E 1994 {\em Phys. Rep.\/} {\bf 244}
77
%
\bibitem{Cop} Copeland E J, Lahiri A and Wands D 1994
\PR\ {\bf D50} 4880
%
\bibitem{louis} Alvarez A, Alvarez-Gaum\'e L and Lozano Y {\em Preprint\/}
CERN-TH.7204/94
%
\end{thebibliography}
\end{document}